\documentclass[aip,jcp,amsmath,amssymb,citeautoscript,longbibliography, reprint]{revtex4-1}

\usepackage[pdftex]{graphics}
\usepackage{subcaption}
\usepackage{tikz}
\usetikzlibrary{calc}
\usetikzlibrary{decorations.pathmorphing, patterns,shapes}
\usepackage{xcolor}

\newif\iffixhighlight
\fixhighlightfalse
\newcommand{\fix}[1]{\iffixhighlight\textcolor{red}{#1}\else#1\fi}

\usepackage{siunitx}
\usepackage{ragged2e}
\usepackage[labelsep=period,format=plain, justification=justified,font=footnotesize]{caption}

\DeclareCaptionJustification{myjust}{\justifying}
\tikzset{snake it/.style={decorate, decoration={snake, amplitude=.5mm, segment length=2mm}}}
\tikzset{zigzag it/.style={decorate, decoration={zigzag, amplitude=.5mm, segment length=2mm}}}

\usepackage{bm}

\DeclareMathAlphabet{\mathsfit}{\encodingdefault}{\sfdefault}{m}{sl}
\SetMathAlphabet{\mathsfit}{bold}{\encodingdefault}{\sfdefault}{bx}{n}

\AtBeginDocument{\def\selectlanguage#1{}}
\DeclareUnicodeCharacter{2009}{\,}  

\begin{document}

\title{Context-Aware Automated NMR Structure Elucidation}

\author{Frank Hu}
\affiliation{Department of Chemistry, Stanford University, Stanford, California, 94305, USA}

\author{Jonathan M. Tubb}
\affiliation{Department of Chemistry, Stanford University, Stanford, California, 94305, USA}

\author{Grant M. Rotskoff}
\email{rotskoff@stanford.edu}
\affiliation{Department of Chemistry, Stanford University, Stanford, California, 94305, USA}

\author{Matthew W. Kanan}
\email{mkanan@stanford.edu}
\affiliation{Department of Chemistry, Stanford University, Stanford, California, 94305, USA}

\author{Thomas E. Markland}
\email{tmarkland@stanford.edu}
\affiliation{Department of Chemistry, Stanford University, Stanford, California, 94305, USA}

\date{\today}

\begin{abstract}
Machine learning approaches are rapidly revolutionizing the ability to analyze and extract structural insights from chemical data, enabling high-throughput chemical discovery. While existing data-driven approaches have shown remarkable promise for predicting chemical structures in the \textit{de novo} setting, incorporating additional chemically informed context offers the opportunity to further improve these models' performance and extend their real-world utility. Here we provide a straightforward protocol to combine artificial-intelligence-based spectroscopic elucidation pipelines with contextual chemical data, e.g., known reactants or a desired product, by using automated retrosynthetic analysis to generate large training sets of molecules and contextual data in the form of synthetic precursors. We demonstrate our approach by showing that our recently introduced NMR structure elucidation framework can be trained to effectively leverage molecular precursor context. We show that when the chemical context provides useful information on the target molecule we obtain dramatically improved accuracy over using NMR spectra alone, obtaining \fix{Top-15} accuracies of up to \fix{95.4\%} for molecules up to 40 heavy atoms spanning the full elemental range of organic chemistry. Furthermore, we demonstrate that our approach is highly robust, such that even when the context provided is uncorrelated with the target, one still obtains accuracies commensurate with those of the original NMR structure elucidation framework. This approach provides an efficient route to improve a wide range of spectroscopic structure elucidation platforms using contextual data.
\end{abstract}


\maketitle

\section{Introduction}
\label{sec:intro}

Recent advances in artificial intelligence have spurred massive growth in research on the use of machine learning techniques to extract rich information on chemical structure, including complete structure elucidation, from spectroscopic data ranging from 1D and 2D NMR\cite{Huang2021,kim_deepsat_2023,alberts_learning_2023,schilter_unveiling_2023,hu_accurate_2024,hu_pushing_2025,jonas_deep_2019,sridharan_deep_2022,devata_deepspinn_2024,jin_nmr-solver_2025,yang_diffnmr_2026} to Mass Spectrometry\cite{bushuiev_massspecgym_2025,bohde_diffms_2025,bohde_frigid_2026}, Infrared\cite{alberts_leveraging_2024}, and UV-vis\cite{tan_transformer_2025}, as well as multimodal models that incorporate combinations of these\cite{xu_spectre_nodate, priessner_enhancing_2024,priessner_enhancing_2025,zipoli_irnmr_2025,alberts_automated_2025}. Many of these models have shown excellent performance in making \textit{de novo} predictions using spectral data alone. However, the combinatorial explosion in chemical space\cite{ertl_cheminformatics_2003,ertl_estimation_2009} for larger molecules, the complexity associated with mixtures of species and impurities, and the noise and finite resolution in real-world experimental conditions remain considerable challenges. At the same time, in experimental settings, structure elucidation is rarely a daunting \textit{de novo} prediction, since contextual information is generally available, ranging from known solvents or reactants to the planned product. Incorporating this contextual information thus offers the opportunity to significantly boost the performance of spectroscopic structure elucidation platforms in the most challenging real-world applications.

A protocol for incorporating additional contextual information in a \textit{de novo} structure elucidation framework should ideally satisfy the following conditions: First, the protocol should be easily applicable to vision-language model style architectures given their proven effectiveness for spectroscopic structure elucidation\cite{ottomano_nmiracle_2026,han_end--end_2026} and since contextual information is best expressed as additional chemical language input. Second, since data scaling is essential for generalizable model performance\cite{kaplan_scaling_2020,henighan_scaling_2020,aghajanyan_scaling_2023,shukor_scaling_2025}, the protocol should allow for efficient training data acquisition. Third, the protocol should not degrade performance in the \textit{de novo} setting and should instead improve accuracy, with worst-case performance reverting to approximately that of the model without any context.

Here, we introduce a protocol for training structure elucidation models to use molecular precursors as context. We demonstrate this approach by applying it to our existing NMRElucidator platform\cite{hu_pushing_2025}, a spectrum-to-structure model exploiting a deep learning architecture inspired by vision-language models. When using only one-dimensional \textsuperscript{1}H and/or \textsuperscript{13}C NMR spectra as the input, we have previously shown that the NMRElucidator can make accurate predictions for molecules with up to 40 non-hydrogen atoms across the full elemental coverage typically encountered in organic chemistry (C, N, O, H, P, S, Si, B, and the halogens), although the accuracy is moderate for molecules at the larger end (Top-15 accuracy of 15\% for 40 heavy atoms) of this size range because of the enormity of the structure space, $\mathcal{O}(10^{37})$ possible molecules. Here, to train NMRElucidator to recognize molecular precursor context, we use AiZynthFinder\cite{genheden_aizynthfinder_2020,saigiridharan_aizynthfinder_2024} to perform retrosynthetic analysis of molecules and sample from the retrosynthetic paths to generate precursors that serve as the prompt for the structure elucidation model. The resulting context-aware NMRElucidator shows an improvement in Top-1 accuracy by up to \fix{46.5} percentage points when provided with high-quality precursor context and minimal ($<$4 percentage points) degradation in accuracy when provided with low-quality context. Our approach is model-agnostic in both directions: we impose no constraints on the structure elucidation model or the retrosynthetic model for generating training precursors, except that the structure elucidation model must be a spectrum-to-structure model. Our protocol therefore provides a route to advance the accuracy of \textit{de novo} spectroscopic structure elucidation models across a wide range of applications.

\section{Results and discussion}

\begin{figure*}
    \centering
    \includegraphics[width=\textwidth]{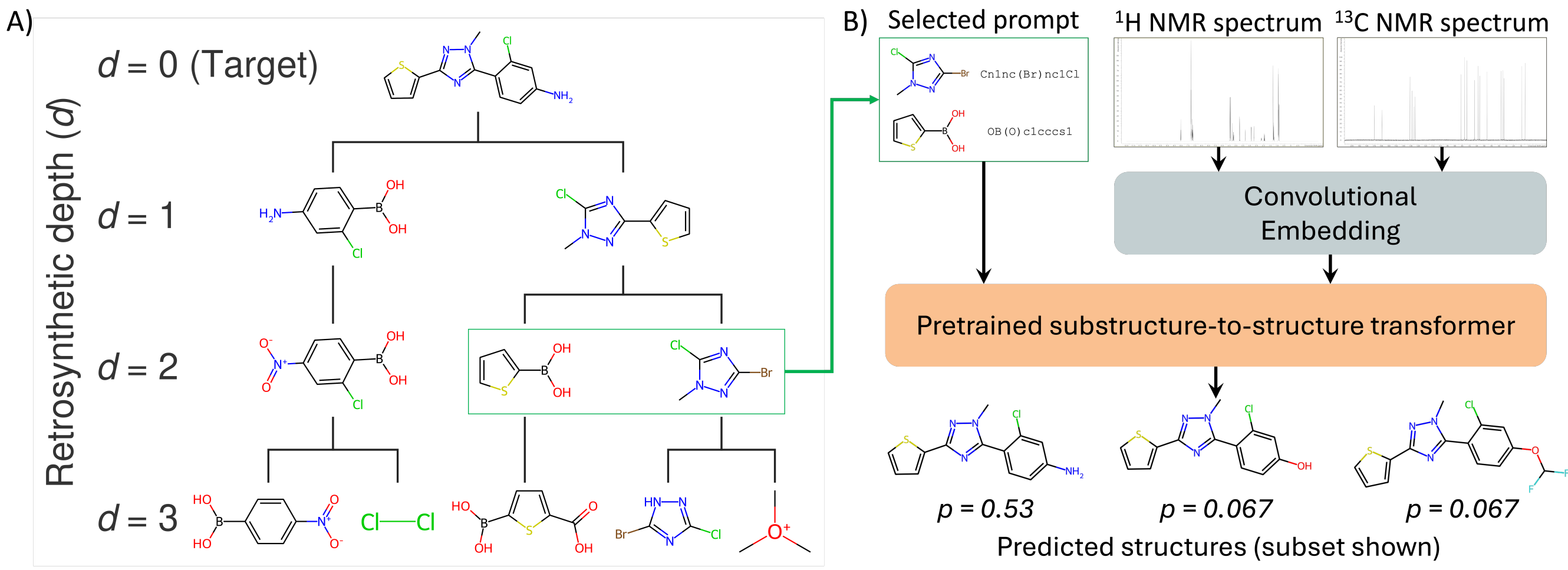}
    \caption{An overview of our context-aware training approach for structure elucidation models using a retrosynthetic tree approach. A) A single retrosynthetic tree for a target molecule. Retrosynthetic trees were generated for molecules to varying retrosynthetic depths, $d$, with the precursors at each level generated via retrosynthetic disconnections of parent molecules at earlier depths. B) Random sampling of the retrosynthetic tree selects out a pair of precursors (green box) to serve as a prompt for the structure elucidation model during training, where it is used alongside the \textsuperscript{1}H~NMR and \textsuperscript{13}C~NMR inputs to elucidate the structure of the corresponding small molecule. A subset of the predicted candidate molecules is shown along with their probabilities $p$ based on a sampling budget of 15 predictions.}
    \label{fig:overview}
\end{figure*} 

The workflow of our prompted training approach is shown in Figure~\ref{fig:overview}. We first construct groups of retrosynthetic trees for each molecule in our dataset using AiZynthExpander\cite{genheden_aizynthfinder_2020,saigiridharan_aizynthfinder_2024} to generate trees of retrosynthetic disconnections one level of retrosynthetic depth at a time. Figure~\ref{fig:overview}A shows a single tree for a target molecule. Starting from the target molecule to generate a single tree, we first generated disconnections to produce possible reactant sets at the first level of retrosynthetic depth $d$. We designate each set of possible reactants as precursors, e.g., those in the green box in Figure~\ref{fig:overview}A. To generate the next level of retrosynthetic depth ($d=2$) we then individually disconnect each reactant molecule to produce another set of precursors. We repeat this process until no further disconnections are predicted or $d=5$. This procedure was used to generate five trees for each molecule. For each molecule in the training, validation, and test sets, we then randomly sample a tree and a set of precursors from the selected tree to serve as the prompt for conditioning the structure elucidation model, as shown by the green box. We chose the retrosynthetic disconnections without constraints on reactant commercial availability or reaction difficulty, ensuring that the precursors used as prompts in both training and testing cover a broad range of potential chemistry, which we expect to make the model more generalizable to unseen reactions \fix{(SI Section~S1\,B)}.

The precursors are provided as a combined SMILES\cite{weininger_smiles_1988} string prompt to the structure elucidation model as shown in Figure~\ref{fig:overview}B. There, it is combined with the NMR input via the cross-attention mechanism in the transformer's\cite{vaswani_attention_2017} decoder stack \fix{(SI Section~S1\,C and SI Figure~S2)}. We chose this approach because it naturally leverages the spectrum-to-structure model's ability to generate conditionally based on textual input. In our previous NMRElucidator model\cite{hu_pushing_2025}, the spectrum input was the embedding of the \textsuperscript{1}H NMR generated by a convolutional neural network combined with the \textsuperscript{13}C NMR spectra, and the language input was just the start token used to begin SMILES structure generation. Here, the language input includes the precursor prompt sampled from the retrosynthetic tree, yielding the \emph{NMR+Precursor} model. The substructure-to-structure transformer in this model utilized our previous generative pretraining strategy where we trained an encoder-decoder transformer model on the task of inverting a molecule's Morgan fingerprint~\cite{rogers_extended-connectivity_2010} back into the structure using $\sim88$M molecules sourced from PubChem~\cite{kim_pubchem_2025}. We then used the pretrained substructure-to-structure transformer weights for end-to-end training of a precursor+spectrum-to-structure encoder-decoder model using both precursors and spectra. The training set consisted of \textsuperscript{13}C and \textsuperscript{1}H NMR spectra for the $\sim2$M molecules ranging from 1 to 40 heavy atoms from our previous work~\cite{hu_pushing_2025} which were augmented by an average of \fix{53} prompts for each molecule ranging from retrosynthetic depths of $d=1$ to $d=5$. The molecules in the train, validation, and test sets were as previously assigned~\cite{hu_pushing_2025}. While we focus on the structure elucidation task in this work, we optimized the model to also predict substructure probabilities for the 2826 substructures we introduced previously using a mixture of the cross entropy loss for the predicted structure and a binary cross entropy loss for the predicted substructures, scalarizing across these two losses with an equivalent weighting factor of 1 (SI Section~S1\,A). End-to-end training of the encoder-decoder transformer multitask model using this procedure yielded the NMR+Precursor model (SI Section~S1). 

To evaluate each model's structure elucidation performance, we use a Top-$k$ random sampling protocol with $k=5$ to sample a set of candidate structures from the model that we then canonicalize before using in any of our comparisons. We contrast this model's performance with two baselines: an \emph{NMR-only} model, i.e., our original NMRElucidator model\cite{hu_pushing_2025}, and a \emph{Precursor-only} model trained on the same training set as the NMR+Precursor model. For the Precursor-only model, we similarly initialize its weights using the pretrained substructure to structure transformer. The former allows us to demonstrate the model's improved performance when precursor prompts are provided as context with the NMR spectra, while the latter lets us compare the prediction accuracy using only precursor information. 

\begin{figure*}
    \centering
    \includegraphics[width=\textwidth]{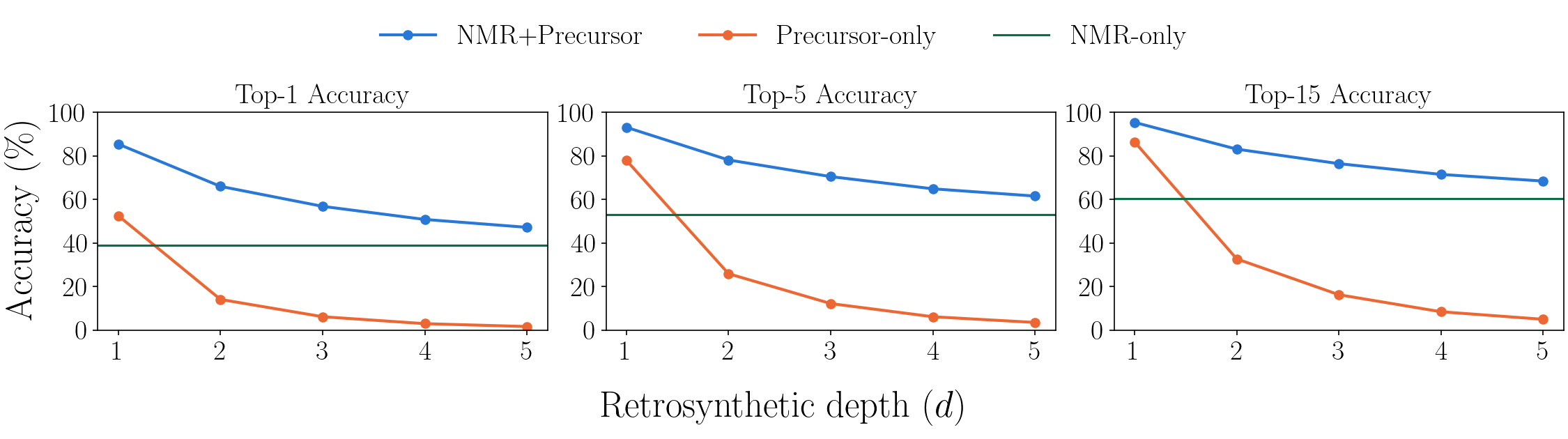}
    \caption{Top-$N$ accuracy ($N=1,5,15$, left to right) as a function of the retrosynthetic depth, $d$, of the precursors provided in the prompt, for the NMR+Precursor (blue), Precursor-only (orange), and NMR-only (green) models. A retrosynthetic depth of $d$ corresponds to precursors obtained after $d$ successive retrosynthetic disconnections of the target. The Top-$N$ accuracy is the percentage of targets for which the correct structure appears among the first $N$ independent predictions. The green horizontal line shows the accuracy of the NMR-only model from our previous work, which does not use a prompt (\fix{38.9\%}, \fix{53.1\%}, and \fix{60.4\%} for $N=1,5,15$, respectively). Accuracies are computed on the full test set of our previous work (201,953 molecules) using, when possible, one randomly selected prompt per molecule at each retrosynthetic depth, excluding prompts in which the target itself appears as a precursor \fix{(SI Table~S2)}.}
    \label{fig:accuracy_vs_depth}
\end{figure*}

Figure~\ref{fig:accuracy_vs_depth} compares the Top-$N$ accuracy ($N=1,5,15$) of the NMR+Precursor \fix{(blue)}, Precursor-only \fix{(orange)}, and NMR-only \fix{(green)} models as a function of the prompt's retrosynthetic depth, $d$. The Top-$N$ accuracy measures the percentage of cases where the correct prediction appears within the first $N$ independent predictions from the model. The test set molecules are the same as those used in our previous work\cite{hu_pushing_2025}, encompassing molecules up to 40 heavy (non-hydrogen) atoms with the full elemental coverage typically encountered in organic chemistry. For each molecule one precursor prompt was randomly sampled at each $d$ \fix{(SI Section~S1\,C)}. 

In all cases, the NMR+Precursor model exceeds the accuracy of both the Precursor-only and the NMR-only models at every $d$. Compared with the NMR-only model, whose accuracy does not depend on $d$ since it does not receive a precursor prompt, the NMR+Precursor model gives the largest accuracy improvement in Top-1 prediction accuracy at $d=1$, rising from \fix{38.9\%} to \fix{85.4\%}---a rise of \fix{46.5} percentage points. This shows that informative precursor context can dramatically improve predictive performance in a single-shot ($N=1$) setting. For Top-15 predictions at $d=1$, the NMR+Precursor model achieves an accuracy of \fix{95.4\%}, which is remarkable given that the test set contains molecules with up to \fix{40} heavy atoms, which are extremely challenging to predict using NMR information alone. Even with precursors from greater depths, the NMR+Precursor model still consistently outperforms the NMR-only model, remaining \fix{8.2}, \fix{8.4}, and \fix{8.0} percentage points above the NMR-only model for $N=1,5,15$ at $d=5$, respectively.

One might expect that the reason for the strong performance of the NMR+Precursor model at low $d$ is due to the high quality of the precursor information, i.e., its structural or synthetic similarity to the target, and that in those cases the result obtained might be no better than just using the precursor information alone. However, the results of the Precursor-only model in Figure~\ref{fig:accuracy_vs_depth} show that the NMR+Precursor model always exceeds its accuracy. In particular, while both models' accuracy decreases monotonically with $d$, the NMR+Precursor model's accuracy decreases less dramatically, dropping by only \fix{27.0} percentage points in Top-15 accuracy from $d=1$ to $d=5$, compared to \fix{81.5} percentage points for the Precursor-only model. This is the ideal behavior because it indicates that while the Precursor-only model makes inaccurate predictions when the context provided by the precursors is far from the target, the NMR+Precursor model is capable of constructively leveraging both modalities of information, with the NMR-only model's performance representing an approximate lower bound on the performance. This failure mode of the Precursor-only model is not alleviated by making more predictions (higher $N$), with a Top-15 accuracy of only \fix{4.9\%} at $d=5$. Importantly, when the NMR+Precursor model is provided with a blank prompt, it recovers the same performance as the NMR-only model, with accuracies of \fix{38.4\%, 52.8\%, and 60.2\%} for $N=1,5,15$, respectively, which are all within \fix{0.6} percentage points of the NMR-only model's accuracies.

\begin{figure}
    \centering
    \includegraphics[width=\columnwidth]{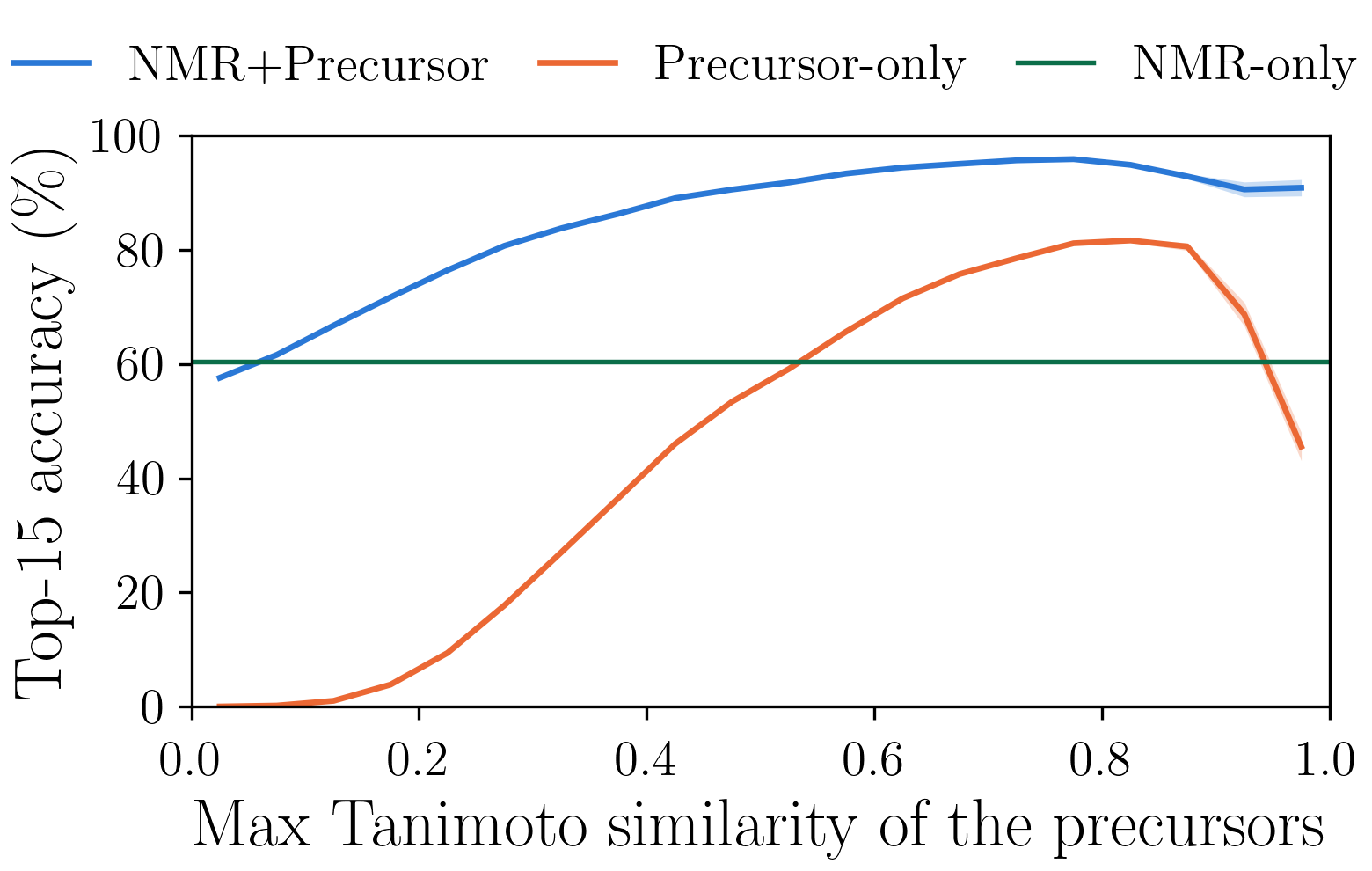}
    \caption{Top-15 accuracy of the NMR+Precursor (blue), Precursor-only (orange), and NMR-only (green) models as a function of the maximum Tanimoto similarity between the precursors in the prompt and the target, pooled over all retrosynthetic depths. Accuracies are computed on the full test set using, when possible, one randomly selected prompt per molecule at each retrosynthetic depth. Prompts in which the target itself appears as a precursor are excluded, as in Figure~\ref{fig:accuracy_vs_depth}, and so are the \fix{132} remaining prompts with a maximum Tanimoto similarity of exactly 1. Shaded bands show the standard error of each accuracy; they are almost indistinguishable from the line width. The green horizontal line shows the Top-15 accuracy of the NMR-only model (\fix{60.4\%}). The Top-1 and Top-5 accuracies are shown in \fix{SI Figure~S5}.}
    \label{fig:acc_vs_tani_and_size}
\end{figure}

To analyze how precursor similarity to the target molecule affects model performance, Figure~\ref{fig:acc_vs_tani_and_size} shows the Top-15 accuracy as a function of the maximum Tanimoto similarity\cite{rogers_computer_1960} of the precursors, defined as the largest Tanimoto similarity when the prompt consists of multiple precursors. As shown in \fix{SI Figure~S4}, the precursor Tanimoto similarity correlates with retrosynthetic depth, decreasing from an average similarity of 0.604 at $d=1$ to 0.160 at $d=5$, but with a broad distribution.

Figure~\ref{fig:acc_vs_tani_and_size} shows that, in cases where the precursor prompt's maximum Tanimoto similarity to the target approaches 0, the Precursor-only model accuracy approaches zero, whereas the NMR+Precursor model gives an accuracy that is close to that of the NMR-only model within a few percentage points, lying only \fix{2.8} percentage points below it when the similarity is below 0.05. Hence, prompts to the NMR+Precursor model that are structurally completely unrelated to the target yield performance close to that of the unprompted (NMR-only) case, indicating little disadvantage to including additional, even unhelpful, context for the model's performance.

As the maximum precursor Tanimoto similarity is increased, as one might expect, the accuracy of both the NMR+Precursor and Precursor-only models initially increases (Figure~\ref{fig:acc_vs_tani_and_size}). However, both models exhibit a maximum accuracy below a Tanimoto similarity of 1: the Top-15 accuracy of the NMR+Precursor model peaks at \fix{95.8\%} for similarities of 0.75--0.80 and dips to \fix{90.6\%} at 0.90--0.95, while that of the Precursor-only model peaks at \fix{81.6\%} for 0.80--0.85 and falls to \fix{45.5\%} at 0.95--1.00. The same effect is also seen in the Top-1 and Top-5 accuracies for both models (\fix{SI Figure~S5}). This corresponds to an interesting case where the precursors in the prompt are structurally very close to the target, differing by a few atoms or bonds, yet both models struggle with the subtle changes needed to reach the target, even with higher sampling budgets. In these cases the models can anchor on the prompt: because a near-identical precursor already reproduces almost the entire target SMILES string, simply returning the precursor is a highly probable output, and the small edit needed to reach the target (e.g., removing a single CH$_2$ unit) cannot be inferred from the precursor alone. Indeed, for the predictions where the prompts have a maximum precursor Tanimoto similarity between 0.95 and 1 (prompts with a similarity of exactly 1 are excluded), \fix{31.0\%} of the Precursor-only model's predictions are exact copies of one of the precursors (identical canonical SMILES) and a further \fix{43.6\%} are structurally closer to one of the precursors than to the target, compared with only \fix{4.8\%} and \fix{7.4\%} respectively for the NMR+Precursor model, consistent with the NMR spectrum providing the evidence needed to make the edit. Some specific examples of this are shown in \fix{SI Section~S2\,C}. However, while in the Precursor-only model this represents a pronounced failure mode, with the accuracy dropping by \fix{31.9}, \fix{39.1}, and \fix{36.1} percentage points for $N=1,5,15$ respectively, for the NMR+Precursor model the decreases from the best accuracies are only \fix{14.0}, \fix{9.2}, and \fix{5.3}. This again highlights the fact that the NMR+Precursor model is extremely capable of constructively leveraging both the NMR and precursor prompt modalities to arrive at better predictions, even in cases where the target and prompts differ by subtle structural differences. 

\begin{figure}
    \centering
    \includegraphics[width=\columnwidth]{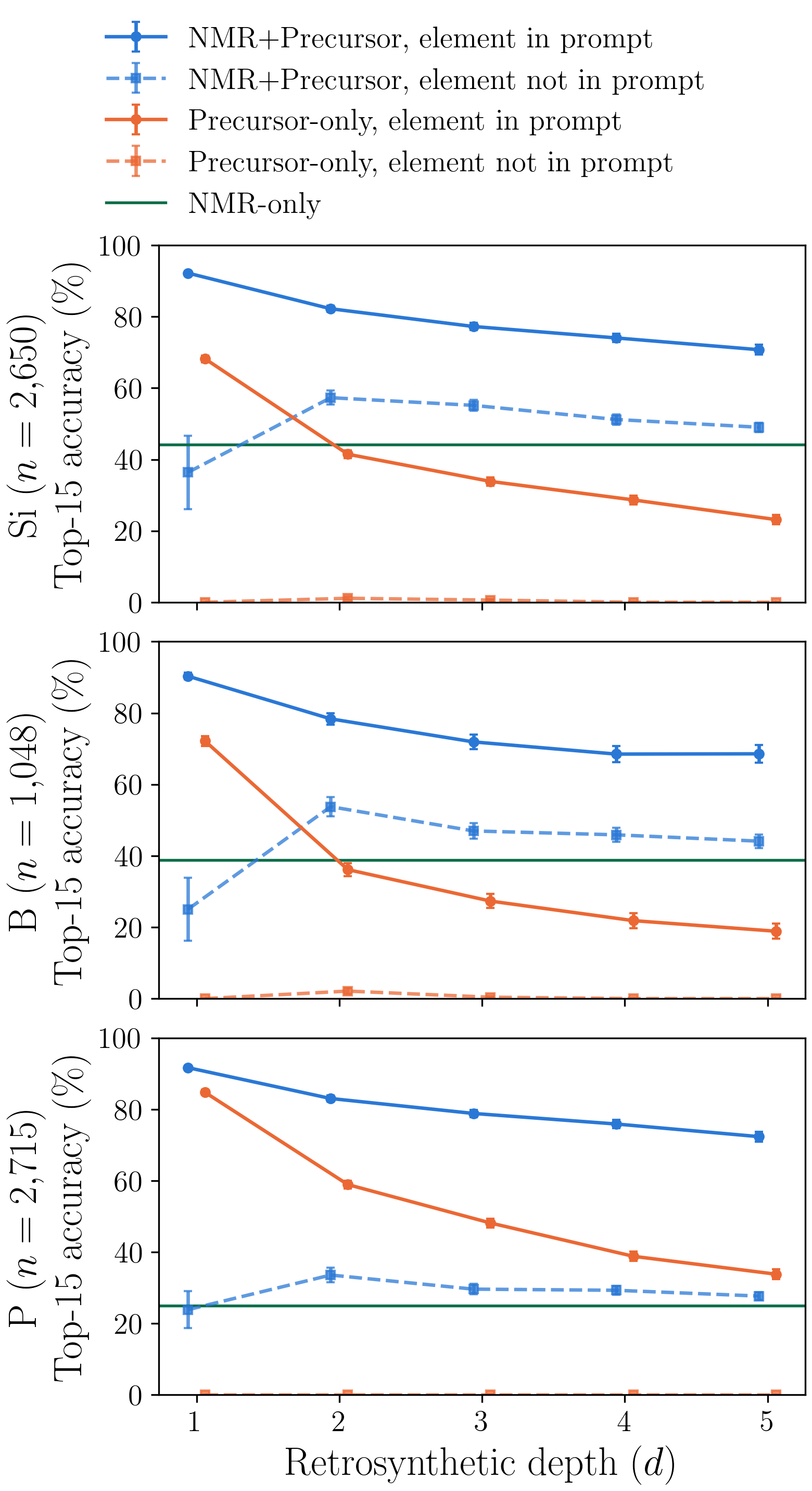}
    \caption{Top-15 accuracy as a function of retrosynthetic depth for target molecules containing the uncommon elements Si (top), B (middle), and P (bottom), for the NMR+Precursor (blue), Precursor-only (orange), and NMR-only (green) models separated into the cases where the prompt does (solid) or does not (dashed) contain the element. Error bars show the standard error of each accuracy, which depends on the number of prompts at that depth and prompt type. The green horizontal lines show the Top-15 accuracy of the NMR-only model on the same molecules (\fix{44.1\%}, \fix{38.7\%}, and \fix{25.0\%} for Si, B, and P, respectively). The Top-1 and Top-5 accuracies, and the numbers of molecules used, are given in \fix{SI Figure~S7}.}
    \label{fig:acc_v_element_presence}
\end{figure}

Elements that are uncommon in organic compounds can pose a significant challenge for NMR-only prediction models since they occur less in the training data and are less frequently predicted by the model. This creates the greatest opportunity for accuracy improvements by providing contextual information. For example, in our previous work we observed that our NMRElucidator model (the NMR-only model in this work) had the lowest accuracy for systems within the test set that contained the elements Si, B, and P. These were some of the most uncommon heteroatoms to appear within the dataset with occurrences in \fix{1.3\%}, \fix{0.5\%}, and \fix{1.3\%} of the \fix{201,953} test molecules, respectively, and had Top-15 accuracies of just \fix{44.1\%}, \fix{38.7\%}, and \fix{25.0\%} compared with \fix{60.4\%} overall. Figure~\ref{fig:acc_v_element_presence} shows the accuracy improvements obtained for these uncommon elements when the prompt either does or does not include them. In all cases we see that inclusion of the element in the prompt dramatically improves the performance across all retrosynthetic depths, $d$, with the NMR+Precursor model reaching Top-15 accuracies of \fix{92.2\%}, \fix{90.3\%}, and \fix{91.7\%} for Si, B, and P at $d=1$---increases of \fix{48.1}, \fix{51.6}, and \fix{66.7} percentage points over the NMR-only model. However, when the uncommon element is not present in the prompt, the accuracy of the NMR+Precursor model resembles that of the NMR-only model, remaining above it for $d \geq 2$ by up to \fix{15.0} percentage points and dropping below it only at $d=1$, where very few prompts lack the element (\fix{22}, \fix{24}, and \fix{67} targets for Si, B, and P, respectively). 

\begin{figure*}[ht]
    \centering
    \includegraphics[width=0.68\textwidth]{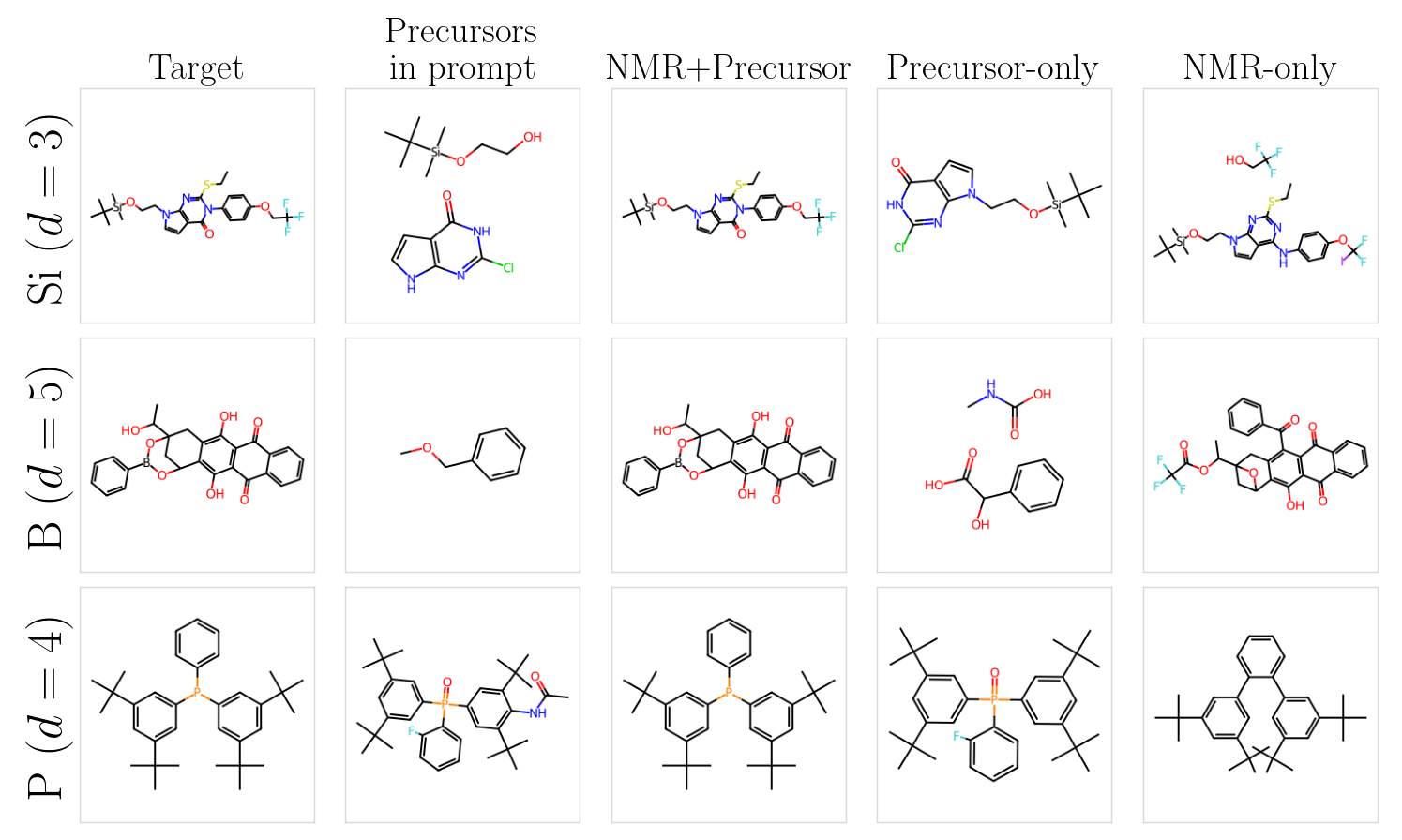}
    \caption{Examples of target molecules containing Si (top), B (middle), and P (bottom) for which only the NMR+Precursor model recovers the correct structure within its Top-15 predictions. Each row, labeled with the target's uncommon element and the retrosynthetic depth $d$ of the prompt, shows, from left to right: the target, the precursors provided in the prompt, and the predictions of the NMR+Precursor, Precursor-only, and NMR-only models. For the NMR+Precursor model, we show the correct structure, which appears among its 15 predictions. The Precursor-only and NMR-only models do not produce the target in any of their 15 predictions, so the prediction with the highest Tanimoto similarity to the target is shown instead. Examples were drawn from the full test set, among single-component targets with at least 30 heavy atoms whose prompt does not contain the target itself and for which the closest prediction of both the Precursor-only and NMR-only models has a Tanimoto similarity to the target below 0.5.}
    \label{fig:Si_B_P_gallery_top_15}
\end{figure*}

Since the Precursor-only model cannot rely on NMR information to obtain a baseline accuracy, Figure~\ref{fig:acc_v_element_presence} shows that its performance for systems containing uncommon elements is entirely dependent on the element being included in the prompt, as the accuracy remains at most \fix{2.1\%} for all $d$ when it is not present. Without an additional modality to determine chemical composition, the Precursor-only model must rely heavily on the prompt, making it essentially impossible for it to determine the correct structure when uncommon elements are present in the target but not in the prompt. Examples of target molecules where the NMR+Precursor model succeeded in obtaining the correct structure while the Precursor-only and NMR-only models failed are shown in Figure~\ref{fig:Si_B_P_gallery_top_15}.

\section{Conclusion}

We have demonstrated a scalable strategy for harnessing contextual information in structure elucidation models. By incorporating retrosynthetic context into our NMR-only model, NMRElucidator, we have created a contextually aware NMR+Precursor model, NMRElucidator+C, which yields significant accuracy improvements, improving Top-1 accuracy by up to 46.5 percentage points. Across all Top-$N$ and retrosynthetic depths, our NMR+Precursor model also outperforms the Precursor-only baseline, showing that it constructively leverages both modalities of information. In particular, we have demonstrated that contextual information provides a particularly significant improvement for predicting molecules with uncommon elements like Si, B, and P: prompts containing the element led to remarkable accuracy improvements of up to \fix{66.7} percentage points over the NMR-only model (Top-15 predictions for P-containing targets with precursors at $d=1$), offering a straightforward way to improve elucidation quality for systems with uncommon heteroatoms. Importantly, all of these improvements in accuracy upon making our NMRElucidator model context-aware were achieved without any significant regression in performance: in cases where the precursor information provided in the prompt is highly uninformative or no prompt is provided at all, the performance of the context-aware \fix{NMRElucidator+C} model matches that of the NMR-only NMRElucidator model in the \textit{de novo} elucidation setting, with the performance improvement scaling directly with the quality of the precursors. Hence, even if a user has very limited or speculative information about the sample they are analyzing, our results suggest that providing this context to our \fix{NMRElucidator+C} model in tandem with their NMR spectrum will at worst provide a prediction of comparable quality to the \textit{de novo} performance of our original NMRElucidator model but, in many cases, provide significant accuracy improvements at no additional computational cost.

While broadly generated retrosynthetic context was used to provide contextual information in this work, our model training strategy is general and thus could be used to incorporate additional forms of context besides reaction precursors, such as reaction conditions, instrument settings, or expected products, into other spectrum-to-structure models, providing a recipe for broadly improving these classes of models and enabling better automated structure elucidation, thus rapidly accelerating the process of chemical synthesis and discovery. 

\section*{Supplementary Material}

\fix{See the supplementary material for details of the structure elucidation model and its training objective, the generation of the retrosynthetic trees (including the AiZynthExpander settings) and of the precursor prompts, the test-set prompt counts, the Precursor-only baselines, and the model optimization protocol, as well as additional results on precursor similarity and size by retrosynthetic depth, accuracy as a function of precursor similarity, examples of the high-similarity accuracy dip, and the Top-1 and Top-5 accuracies for targets containing Si, B, and P.}

\section*{Acknowledgments}

F.H. acknowledges support from a Stauffer fellowship. Some of the computing for this project was performed on the Sherlock cluster. We would like to thank Stanford University and the Stanford Research Computing Center for providing computational resources and support that contributed to these research results. J.M.T. acknowledges support from the Stanford Undergraduate Research Fellowship in Chemistry. We thank Hongyi Zhang for helpful discussions on NMR spectroscopy.





\section*{References}
\bibliography{references}

\end{document}


\title{Supplementary Information: Context-Aware Automated NMR Structure Elucidation}
\maketitle


\section{Methods}
\label{SI_sec:si_methods}

Section~\ref{SI_subsec:struct-elucidate} describes the automated structure elucidation framework that we use and our protocol for training this model without precursor information. Section~\ref{SI_subsec:retrosynthesis} then describes how the retrosynthetic trees are generated, and Section~\ref{SI_subsec:retro-and-struct-elucidate} how prompts are built from them for input to the structure elucidation model. We additionally include details regarding how we trained our baseline Precursor-only model (Section~\ref{SI_subsec:baseline-models}), and details on model optimization (Section~\ref{SI_subsec:model-optimization}). 

\subsection{Automated structure elucidation}\label{SI_subsec:struct-elucidate}

In this work, we focus on \textit{de novo} structure elucidation from 1D \textsuperscript{1}H and \textsuperscript{13}C NMR, using our recently introduced NMRElucidator\cite{hu_pushing_2025} multitask model. The model takes in \textsuperscript{1}H NMR, represented as a raw vector of intensities, and the list of \textsuperscript{13}C NMR shifts and produces a prediction for the molecular structure represented as a SMILES string as well as a prediction of the substructures contained within the molecule. We chose this model because of its vision-language architecture, relatively low compute requirements, and competitive performance. The methods we develop below are in principle agnostic to the model so long as it is a spectrum-to-structure model with a vision-language style architecture.

Consider an \textsuperscript{1}H NMR spectrum represented as a vector of $n_{\text{H}}$ normalized intensities $\xb_{\text{H}} \in [0,1]^{n_{\text{H}}}$, a vector of $n$ \textsuperscript{13}C NMR chemical shifts $\xb_{\text{C}} \in \sR^{n}$, and a vector of $m$ substructure probabilities $\yb_{\text{sub}} \in [0,1]^m$, all of which are associated with a specific molecule represented by its SMILES string $\yb_{\text{smi}}$. Let $\gM_{\theta}$ denote the NMRElucidator model with parameters $\theta$. The model operates by taking the concatenated input vector $\xb = [\xb_{\text{H}}, \xb_{\text{C}}]$, and producing a prediction for both the SMILES string of the molecule and its substructure profile, $\hat{\yb}_{\text{smi}}, \hat{\yb}_{\text{sub}} = \gM_{\theta}(\xb)$. The prediction for the SMILES string proceeds via autoregressive decoding as in most standard language models, and is optimized using a cross entropy loss of the form

\begin{equation}
    \mathcal{L}_{\text{CE}}(\theta)=-\frac{1}{L}\sum_{i=1}^L\log \gM_{\theta}(\yb_{\text{smi}}^i | \yb_{\text{smi}}^{<i},\xb),
    \label{SI_eq:CE_loss}
\end{equation}

where $L$ is the length of the SMILES string in terms of tokens and $\yb_{\text{smi}}^{<i}$ denotes all tokens before position $i$. The prediction for the substructure profile is a single vector $\hat{\yb}_{\text{sub}}$ where each element $y_i\in\hat{\yb}_{\text{sub}}$ is the probability that the $i$th substructure is present. We optimize this using a binary cross entropy loss of the form

\begin{equation}
    \mathcal{L}_{\text{BCE}}(\theta) = -\frac{1}{m}\sum_{i=1}^m \left[\yb_{\text{sub}}^i \log \hat{\yb}_{\text{sub}}^i + (1 - \yb_{\text{sub}}^i)\log (1-\hat{\yb}_{\text{sub}}^i)\right].
\end{equation}

The model is jointly optimized by scalarizing across the two terms with weighting factors $\alpha$ and $\beta$, with $\mathcal{L}(\theta)=\alpha\mathcal{L}_{\text{CE}}(\theta)+\beta\mathcal{L}_{\text{BCE}}(\theta)$. For simplicity, we take $\alpha = \beta = 1$, ensuring an equal weighting of the two loss terms. Because of the structure of the NMRElucidator model where the substructure and structure prediction branches are separated within the model architecture, we primarily focus on context injection for the structure elucidation branch of the model. We use the same set of 2826 predefined substructures as in previous work. Our focus in this work is on the structure elucidation portion of the model, but we maintain the substructure portion of the loss. For all of our model trainings, we start from the pretrained substructure-to-structure fingerprint transformer of our previous work.

\subsection{Retrosynthetic trees for molecules}\label{SI_subsec:retrosynthesis}

Retrosynthesis involves the analysis of molecular structures in terms of synthetically accessible disconnections, with the aim of reducing a complex target product into simpler, commercially accessible building blocks. This enables the forward design of synthetic routes, converting precursors into products. While obtaining a single retrosynthetic route is useful, an important aspect to our approach is having a scalable strategy for generating retrosynthetic routes that cover a wide range of possible chemistry, since we want to train a structure elucidation model that can leverage precursor information across a wide range of scenarios. Consequently, rather than generating a single retrosynthetic route for a molecule, we wish to generate a retrosynthetic \textit{tree} such that we can obtain many different retrosynthetic routes per tree.

Starting with a molecule $M_0$, we construct a retrosynthetic tree $\gT$ with $M_0$ as the root node, expanding at the first level using a conditional policy $\pi_E(\cdot|M_0)$ to obtain several suitable one-step-back precursor sets for $M_0$, from which a single set is chosen for inclusion in the tree. We then recursively apply the policy $\pi_E$ at every subsequent level to each node, repeating the sampling process and further expanding the tree until no further disconnections are predicted or a maximum depth of 5 is reached. By traversing the tree, we can obtain multiple sets of precursors that lead to a given product, allowing us to control the depth from which the precursors originate. Having access to precursors generated at different depths within the retrosynthetic tree is useful because it allows for easily assessing the utility of precursor information that is structurally further from or closer to the desired product, as moving further down the tree leads to molecules that are more structurally dissimilar. Furthermore, by using a chemically informed policy function $\pi_E$ for the expansion, we are guaranteed that every expansion of the tree proceeds along some chemically possible transformation.

For the tree expansion policy, we choose to use AiZynthExpander, a neural network guided approach that is provided as part of the AiZynthFinder package\cite{genheden_aizynthfinder_2020,saigiridharan_aizynthfinder_2024}. We additionally do not filter the allowable types of transformations based on factors such as feasibility or reagent cost: the feasibility filter model only records a score for each predicted reaction, and no reactions are removed on that basis, as we want the widest possible coverage of the chemistries representable under the expansion policy. Much like the choice of structure elucidation model, we note that our context-aware training approach is agnostic to the choice of the retrosynthetic tree generator, with many other template-based and deep learning approaches available. We choose AiZynthExpander because it is open source and highly efficient when generating trees for complex molecules, taking on the order of 0.1--0.2 seconds per molecule. 

\begin{figure}[!htbp]
    \centering
    \includegraphics[width=\textwidth]{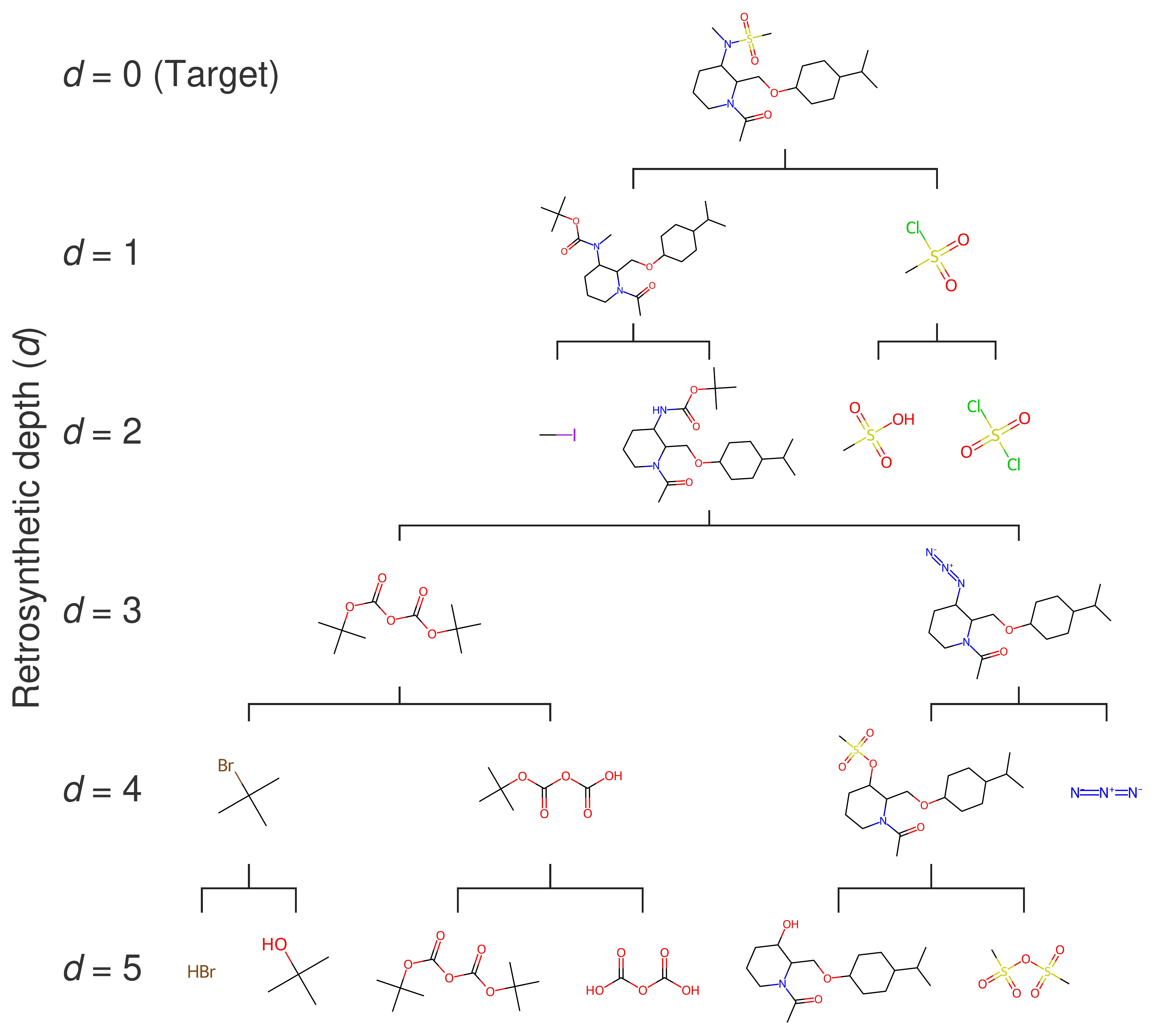}
    \caption{An example retrosynthetic tree generated by AiZynthExpander for a test-set molecule (top), one of the five trees generated for this molecule, expanded to the maximum retrosynthetic depth of $d=5$. Each row shows the precursors at retrosynthetic depth $d$, and the lines connect each molecule to the precursors obtained from its disconnection; the children of a molecule together form one possible prompt.}
    \label{SI_fig:example_tree}
\end{figure}

For generating the dataset of prompts, we begin with our full dataset from the original NMRElucidator training, which consists of 1{,}999{,}897 molecules partitioned into 1{,}595{,}793 for training, 202{,}151 for validation, and 201{,}953 for testing. All molecules are provided alongside their \textsuperscript{1}H and \textsuperscript{13}C NMR spectra. For each of these molecules, we use AiZynthExpander to generate five different retrosynthetic trees, each beginning from a distinct first disconnection of the target, terminating trees at a maximum retrosynthetic depth of $d=5$ or until no further disconnections are predicted. An example retrosynthetic tree is shown in Figure~\ref{SI_fig:example_tree}. All of the relevant settings for generating the trees can be found in Table~\ref{SI_tab:aizynth_settings}.

\begin{table}[h]
    \centering
    \renewcommand{\arraystretch}{1.2}
    \caption{Settings used to generate the retrosynthetic trees with AiZynthExpander. Settings not listed are AiZynthFinder defaults.}
    \label{SI_tab:aizynth_settings}
    \begin{tabular}{lp{0.5\textwidth}}
        \hline
        Setting & Value \\
        \hline
        Software & AiZynthFinder 4.4.1 (\texttt{AiZynthExpander}) \\
        Expansion policy & USPTO template-based neural network (\texttt{uspto\_model.onnx}, \texttt{uspto\_templates.csv.gz}) \\
        Template cutoffs & cumulative probability 0.995, at most 50 templates (\texttt{cutoff\_cumulative}, \texttt{cutoff\_number}) \\
        Reactant sets returned per expansion (\texttt{return\_n}) & 50 \\
        Feasibility (filter) model & USPTO (\texttt{uspto\_filter\_model.onnx}); scores recorded only, no reactions removed \\
        Selection at each node & uniformly at random among the returned reactant sets \\
        Trees per molecule & 5, each from a distinct first disconnection \\
        Maximum depth & 5 \\
        Random seed & 42 \\
        \hline
    \end{tabular}
\end{table}

\FloatBarrier

\subsection{Retrosynthetic prompting as context}\label{SI_subsec:retro-and-struct-elucidate}

We now turn our attention to constructing context from the retrosynthetic tree. During training for a given molecule and a randomly sampled retrosynthetic tree $\gT$ of depth $K$ from its set of five, we first randomly sample a depth $k < K$ and traverse the tree to that level, aggregating all of the nodes at that level of the tree that have children (i.e., that were expanded). We then randomly sample a node from that collection and concatenate the SMILES strings of its children to use as the prompt with a special separator token specified within the model alphabet. Notably, while the depth of the parent node is selected such that $k \in [0, K-1]$, the retrosynthetic depth of the precursors themselves is offset by one, hence $d \in [1, K]$ as is shown for all the results in the main text. 

The inclusion of this prompt slightly modifies the structure elucidation training objective presented in Equation~\ref{SI_eq:CE_loss} as follows:

\begin{equation}
    \mathcal{L}_{\text{CE}}(\theta)=-\frac{1}{L}\sum_{i=1}^L\log \gM_{\theta}(\yb_{\text{smi}}^i | \yb_{\text{smi}}^{<i}, \yb_{\text{precursor}}, \xb),
    \label{SI_eq:CE_loss_prompt}
\end{equation}

where we now have the additional term $\yb_{\text{precursor}}$ which represents the tokenized prompt. While the entire sequence input is a concatenation of $\yb_{\text{precursor}}$ and $\yb_{\text{smi}}$, the loss is only computed over the completion portion, ensuring that the model is only evaluated on its response as opposed to also having the loss computed over the prompt. 

For evaluation using the retrosynthetic trees, we sample one prompt randomly at each depth $d$ for each molecule from its retrosynthetic trees when possible, as not all retrosynthetic expansions for molecules are successful. In evaluation, we exclude prompts that include the target as part of the prompt, which leads to further data attrition. For the actual number of molecules and prompts used for evaluation, refer to Table~\ref{SI_tab:prompt_counts}.

\begin{table}[h]
    \centering
    \setlength{\tabcolsep}{10pt}  
    \renewcommand{\arraystretch}{1.2}
    \caption{Number of test-set prompts at each retrosynthetic depth $d$ before and after excluding prompts in which the target molecule itself appears among the precursors. Each of the 201{,}953 test molecules contributes at most one randomly selected prompt per depth; a molecule has no prompt at a depth if none of its retrosynthetic trees reaches that depth (577 molecules have no usable tree at all). The NMR+Precursor and Precursor-only models are evaluated on identical prompts.}
    \label{SI_tab:prompt_counts}
    \begin{tabular}{crrrr}
        \hline
        $d$ & \multicolumn{1}{c}{Molecules with a prompt} & \multicolumn{1}{c}{Target among precursors} & \multicolumn{1}{c}{Prompts used} & \multicolumn{1}{c}{\% of test set} \\
        \hline
        1 & 201{,}376 & 55 (0.03\%) & 201{,}321 & 99.69 \\
        2 & 201{,}302 & 8{,}057 (4.00\%) & 193{,}245 & 95.69 \\
        3 & 201{,}280 & 708 (0.35\%) & 200{,}572 & 99.32 \\
        4 & 201{,}271 & 976 (0.48\%) & 200{,}295 & 99.18 \\
        5 & 201{,}266 & 183 (0.09\%) & 201{,}083 & 99.57 \\
        \hline
        Total & 1{,}006{,}495 & 9{,}979 (0.99\%) & 996{,}516 & \\
        \hline
    \end{tabular}
\end{table}

\begin{figure}[!htbp]
    \centering
    \includegraphics[width=0.85\textwidth]{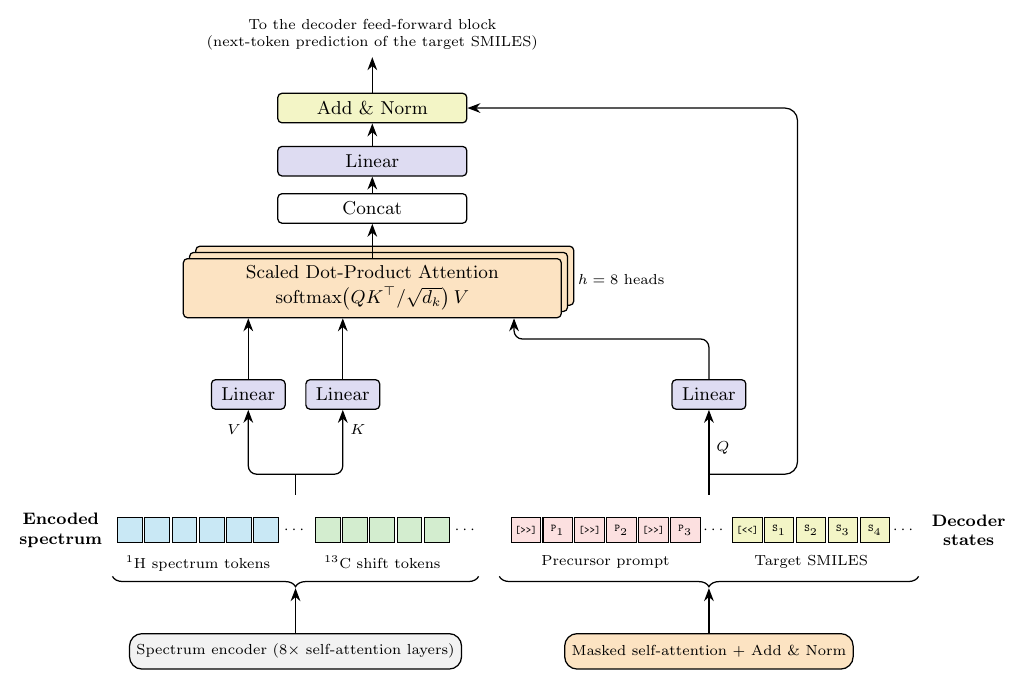}
    \caption{Cross-attention between the NMR spectra and the precursor prompt, as performed in each of the eight decoder layers of the NMR+Precursor structure model. Keys ($K$) and values ($V$) are linear projections of the output of the spectrum encoder, whose input sequence consists of the \textsuperscript{1}H NMR spectrum tokens produced by the convolutional embedding followed by one token per \textsuperscript{13}C NMR chemical shift. Queries ($Q$) are linear projections of the decoder states of the precursor prompt and target SMILES tokens after masked self-attention. Attention is computed in $h=8$ heads, whose outputs are concatenated, linearly projected, and added to the decoder states (Add \& Norm) before the decoder's feed-forward block. The cross-attention layers are the only place in the model where the precursor prompt and the NMR spectra interact.}
    \label{SI_fig:cross_attention}
\end{figure}

Architecturally, the embedding from the tokenized sequence of precursors is only combined with the concatenated representation of the spectra $\xb = [\xb_{\text{H}}, \xb_{\text{C}}]$ within the cross-attention layers of the decoder, where the latent representation of the spectrum embeddings generated by the encoder network is combined with the processed embeddings of the precursor and SMILES\cite{weininger_smiles_1988} sequence after the autoregressive masked attention, as shown in Figure~\ref{SI_fig:cross_attention}. The structure follows closely that of the original transformer introduced in Vaswani et al.\cite{vaswani_attention_2017} 

\FloatBarrier

\subsection{Precursor-only baselines}\label{SI_subsec:baseline-models}

For our baseline Precursor-only model, we use an encoder-decoder style architecture by isolating the encoder-decoder transformer stack from the substructure-to-structure model pretrained in our previous work. Because that model is pretrained on the task of converting Morgan fingerprints\cite{rogers_extended-connectivity_2010} to SMILES strings, we maintain all of the weights but replace the embedding layer so that it is amenable to the string-based representation of the concatenated precursors and SMILES strings. 

\begin{figure}[!htbp]
    \centering
    \includegraphics[width=\textwidth]{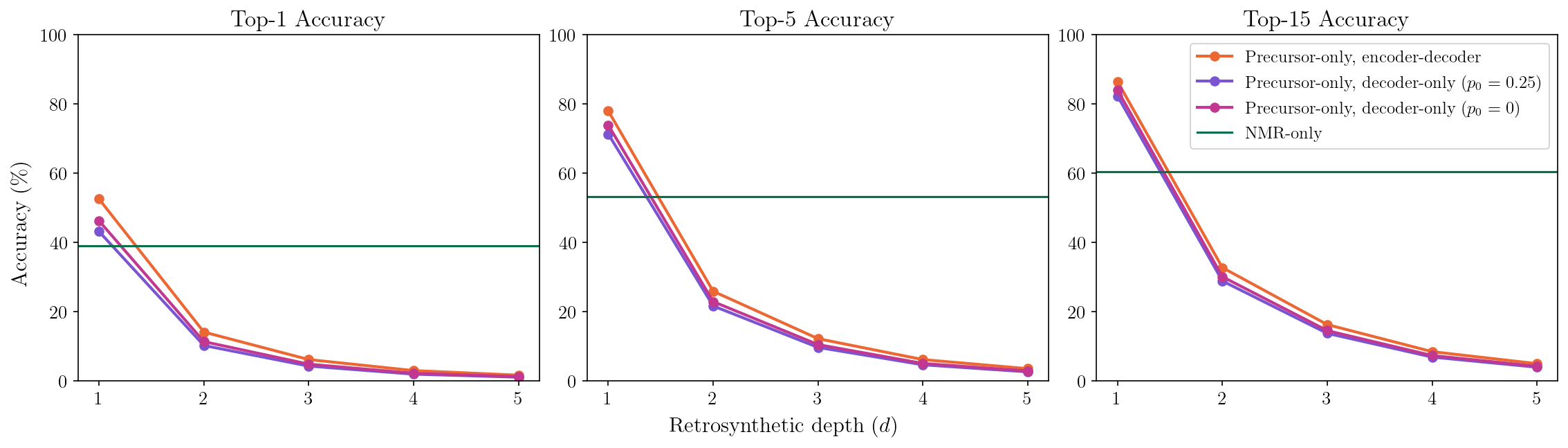}
    \caption{Top-$N$ accuracy ($N=1,5,15$, left to right) as a function of the retrosynthetic depth $d$ of the precursors in the prompt for the Precursor-only baselines: the encoder-decoder model used for all comparisons in the main text (orange), and the decoder-only model trained with a probability $p_0 = 0.25$ (purple) or $p_0 = 0$ (magenta) of replacing the training prompt by a blank prompt. The green horizontal lines show the accuracy of the NMR-only model (38.9\%, 53.1\%, and 60.4\% for $N=1,5,15$, respectively). Accuracies are computed on the full test set, using one randomly selected prompt per molecule at each retrosynthetic depth when possible and excluding prompts in which the target itself appears as a precursor.}
    \label{SI_fig:baselines}
\end{figure}

In addition to the encoder-decoder Precursor-only model, we tested additional variations that correspond to a decoder-only architecture, where we just retained the decoder stack of the pretrained transformer. This required few modifications to the embedding layer since the decoder stack of the full encoder-decoder pretrained transformer is already accustomed to representing and generating SMILES strings. We find that the decoder-only model is less accurate than the encoder-decoder model at every retrosynthetic depth, by up to 9.4 percentage points in Top-1 accuracy and 4.2 percentage points in Top-15 accuracy (both at $d=1$), and so choose the encoder-decoder Precursor-only model variation for all the comparisons presented in the main text. For a comparison of all the baseline performances, see Figure~\ref{SI_fig:baselines}.

\FloatBarrier

\subsection{Model optimization protocol}\label{SI_subsec:model-optimization}

All models were trained with PyTorch Lightning\cite{paszke_pytorch_2019} on a single NVIDIA H100 GPU, using a batch size of 128, a constant learning rate of $10^{-5}$ with no learning-rate schedule, and a fixed random seed. Each time a training molecule is drawn, a prompt is sampled at random from its retrosynthetic tree and, with probability $p_0$, replaced by a blank prompt containing no precursors. For every model, the cross-entropy loss over the predicted SMILES is evaluated only over the tokens of the target SMILES string, masking the tokens of the precursor prompt that precede it.

The NMR+Precursor model was optimized with AdamW (moment coefficients 0.9 and 0.999, $\epsilon = 10^{-8}$, weight decay 0.01) in 32-bit precision, with gradients clipped to a maximum norm of 1.0, on the equally weighted sum of the structure and substructure losses described above, and with $p_0 = 0.25$. The Precursor-only models were optimized with Adam\cite{kingma_adam_2017} (moment coefficients 0.9 and 0.999, $\epsilon = 10^{-8}$, no weight decay) in bf16 mixed precision without gradient clipping, on the structure loss alone. The encoder-decoder Precursor-only model used for the comparisons in the main text was trained with $p_0 = 0$, and the decoder-only variant was trained with both $p_0 = 0.25$ and $p_0 = 0$. All models were allowed up to 1200 epochs and were stopped after 84 (NMR+Precursor), 112 (encoder-decoder Precursor-only), and 180 (both decoder-only Precursor-only) epochs. In each case we use the checkpoint with the lowest validation loss, as summarized in Table~\ref{SI_tab:optimization}.

\begin{table}[h]
    \centering
    \caption{Optimization settings for each model. All models use a batch size of 128, a constant learning rate of $10^{-5}$, and a single NVIDIA H100 GPU. $p_0$ is the probability that a training prompt is replaced by a blank prompt. The best epoch is the epoch with the lowest validation loss, whose checkpoint is used for evaluation.}
    \label{SI_tab:optimization}
    \begin{tabular}{lcccccc}
        \hline
        Model & Optimizer & Precision & Gradient clipping & $p_0$ & Epochs trained & Best epoch \\
        \hline
        NMR+Precursor & AdamW & FP32 & norm 1.0 & 0.25 & 84 & 50 \\
        Precursor-only, encoder-decoder & Adam & bf16 mixed & none & 0 & 112 & 112 \\
        Precursor-only, decoder-only & Adam & bf16 mixed & none & 0.25 & 180 & 147 \\
        Precursor-only, decoder-only & Adam & bf16 mixed & none & 0 & 180 & 171 \\
        \hline
    \end{tabular}
\end{table}

\section{Additional results}\label{SI_sec:addn-results}

\subsection{Precursor similarity and size by retrosynthetic depth}\label{SI_subsec:precursor-stats}

\begin{figure}[!htbp]
    \centering
    \includegraphics[width=\textwidth]{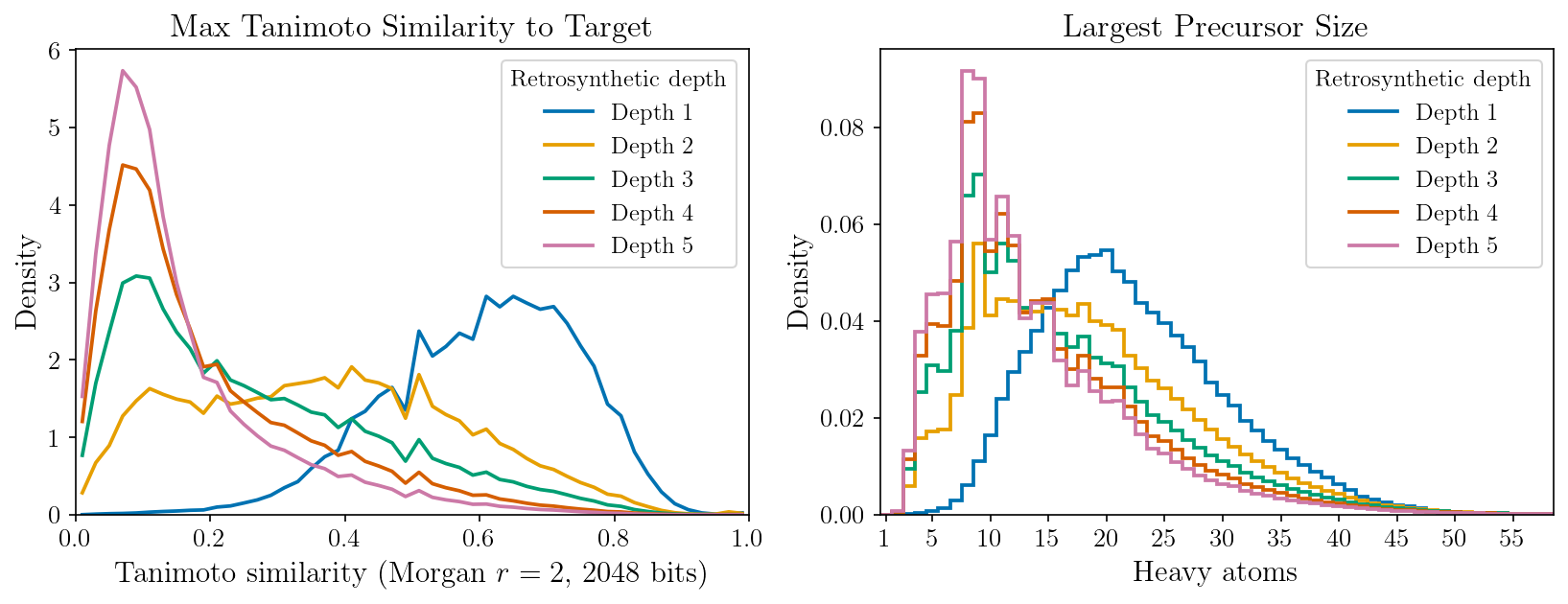}
    \caption{Distributions of the maximum Tanimoto similarity to the target (left) and of the size of the largest precursor in heavy atoms (right), taken over the precursors in each prompt and separated by retrosynthetic depth. Tanimoto similarities are computed from Morgan fingerprints (radius 2, 2048 bits). The distributions are computed over all prompts in the retrosynthetic trees of the full dataset of $\sim$2 million molecules; prompts in which the target itself appears as a precursor are excluded from the Tanimoto similarity distributions.}
    \label{SI_fig:tree_diagnostics}
\end{figure}

Figure~\ref{SI_fig:tree_diagnostics} shows how the maximum Tanimoto similarity of the precursors to the target and the size of the largest precursor change with retrosynthetic depth across the retrosynthetic trees.

\FloatBarrier  

\subsection{Accuracy as a function of precursor similarity}\label{SI_subsec:acc-vs-similarity}

\begin{figure}[!htbp]
    \centering
    \includegraphics[width=\textwidth]{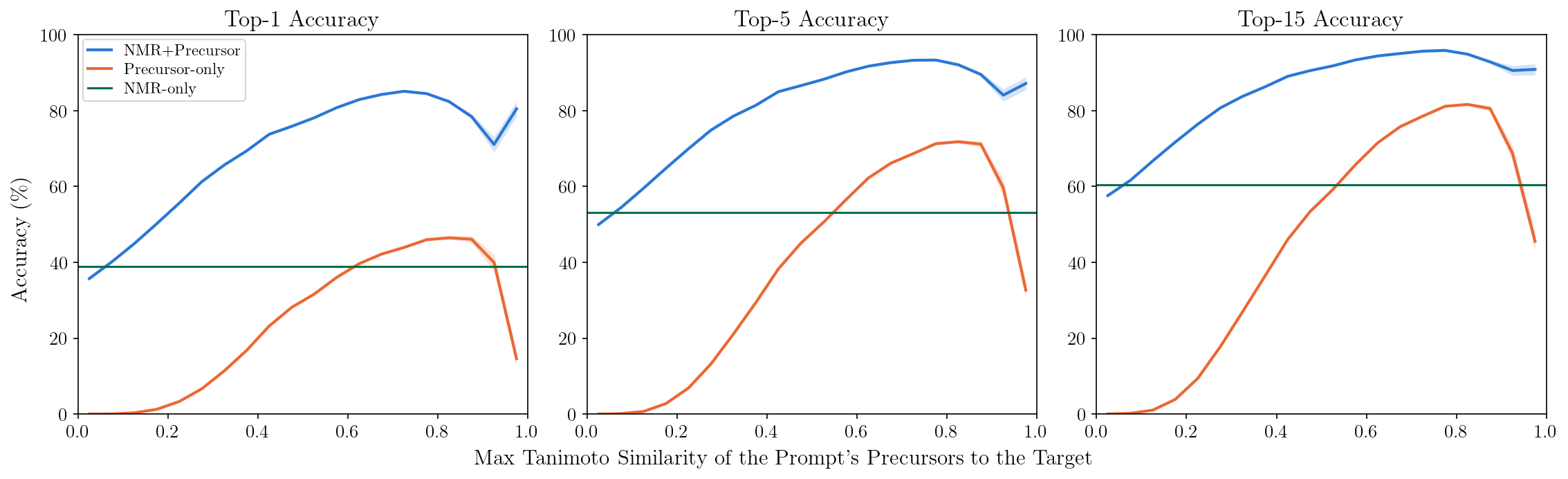}
    \caption{Top-$N$ accuracy ($N=1,5,15$, left to right) of the NMR+Precursor (blue), Precursor-only (orange), and NMR-only (green) models as a function of the maximum Tanimoto similarity between the precursors in the prompt and the target, pooled over all retrosynthetic depths. Accuracies are computed on the full test set using, when possible, one randomly selected prompt per molecule at each retrosynthetic depth. Prompts in which the target itself appears as a precursor are excluded, and so are the 132 remaining prompts with a maximum Tanimoto similarity of exactly 1. Shaded bands show the standard error of each accuracy. The green horizontal lines show the accuracy of the NMR-only model (38.9\%, 53.1\%, and 60.4\% for $N=1,5,15$, respectively). The Top-15 panel is shown in the main text (Figure~3).}
    \label{SI_fig:acc_vs_tani_and_size_full}
\end{figure}

Figure~\ref{SI_fig:acc_vs_tani_and_size_full} shows the Top-1, Top-5, and Top-15 accuracies as a function of the maximum precursor Tanimoto similarity.

\FloatBarrier  

\subsection{Examples of the high-similarity dip}\label{SI_subsec:dip-examples}

\begin{figure}[!htbp]
    \centering
    \includegraphics[width=0.8\textwidth]{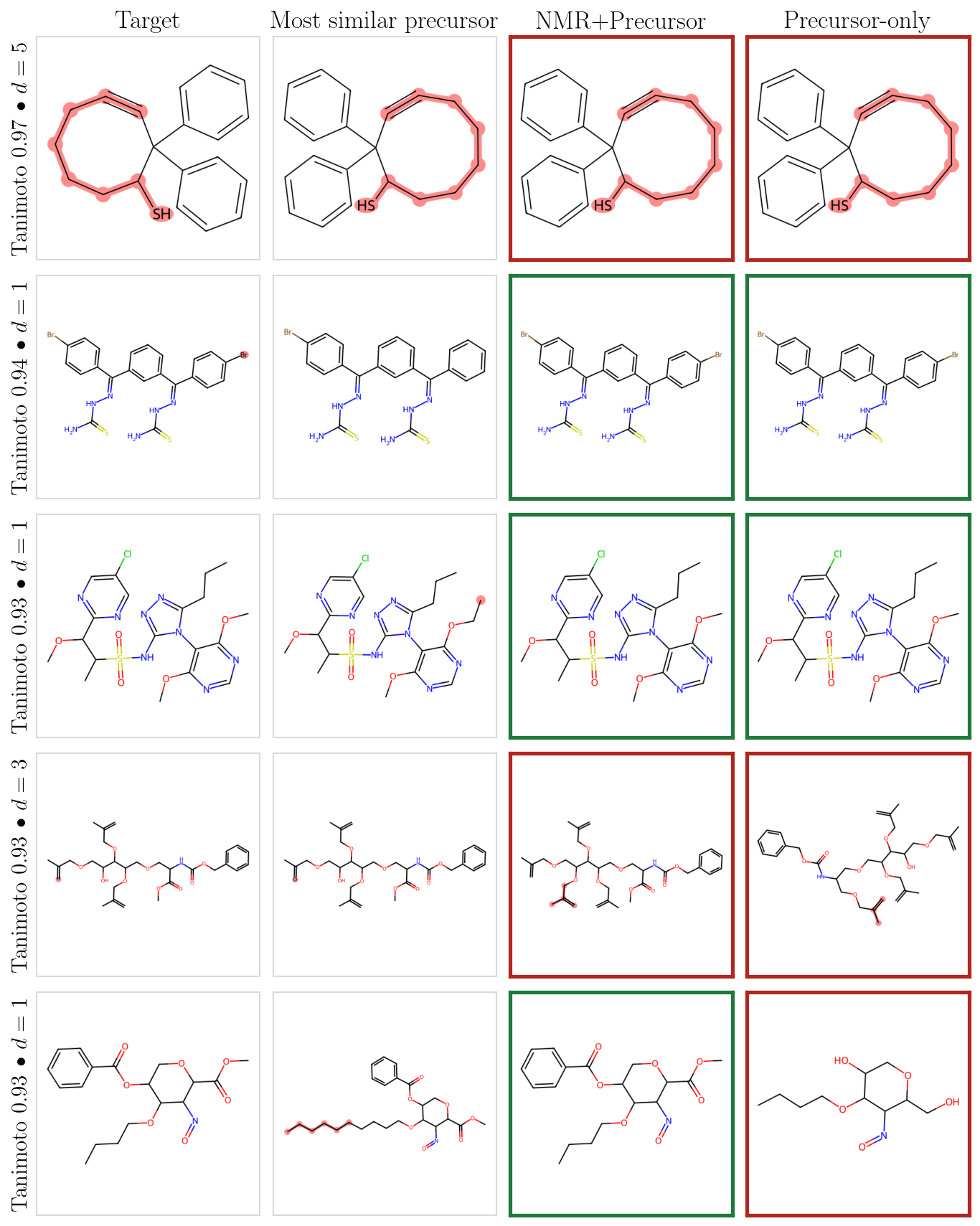}
    \caption{Five randomly selected prompts with a maximum precursor Tanimoto similarity between 0.90 and 1.0, drawn from the full test set with prompts in which the target itself appears as a precursor excluded, and excluding prompts whose most similar precursor differs from the target only in its salt form or counterions. Each row shows, from left to right, the target (labeled with the maximum Tanimoto similarity and the retrosynthetic depth $d$ of the prompt), the precursor in the prompt most similar to the target, and the first (Top-1) prediction of the NMR+Precursor and Precursor-only models. Atoms highlighted in red lie outside the maximum common substructure with the target (for the target, with the most similar precursor). Prediction boxes are outlined in green if the prediction is the target and in red otherwise.}
    \label{SI_fig:dip_examples}
\end{figure}

Figure~\ref{SI_fig:dip_examples} shows examples of prompts from the high-similarity region where the accuracy of both models dips (maximum precursor Tanimoto similarity between 0.90 and 1.0). In these prompts the most similar precursor is a near-copy of the target: in the 0.95--1.00 bin, 78.2\% of the most similar precursors differ from the target only in their salt form or counterions, and in the 0.90--0.95 bin, 54.9\% are homologs of the target, differing by one or more CH$_2$ units. In the 0.95--1.00 bin, the first prediction of the Precursor-only model is a copy of the most similar precursor for 31.9\% of prompts, compared with 4.2\% for the NMR+Precursor model.

\FloatBarrier  

\subsection{Targets containing Si, B, and P}\label{SI_subsec:rare-elements}

\begin{figure}[!htbp]
    \centering
    \includegraphics[width=\textwidth]{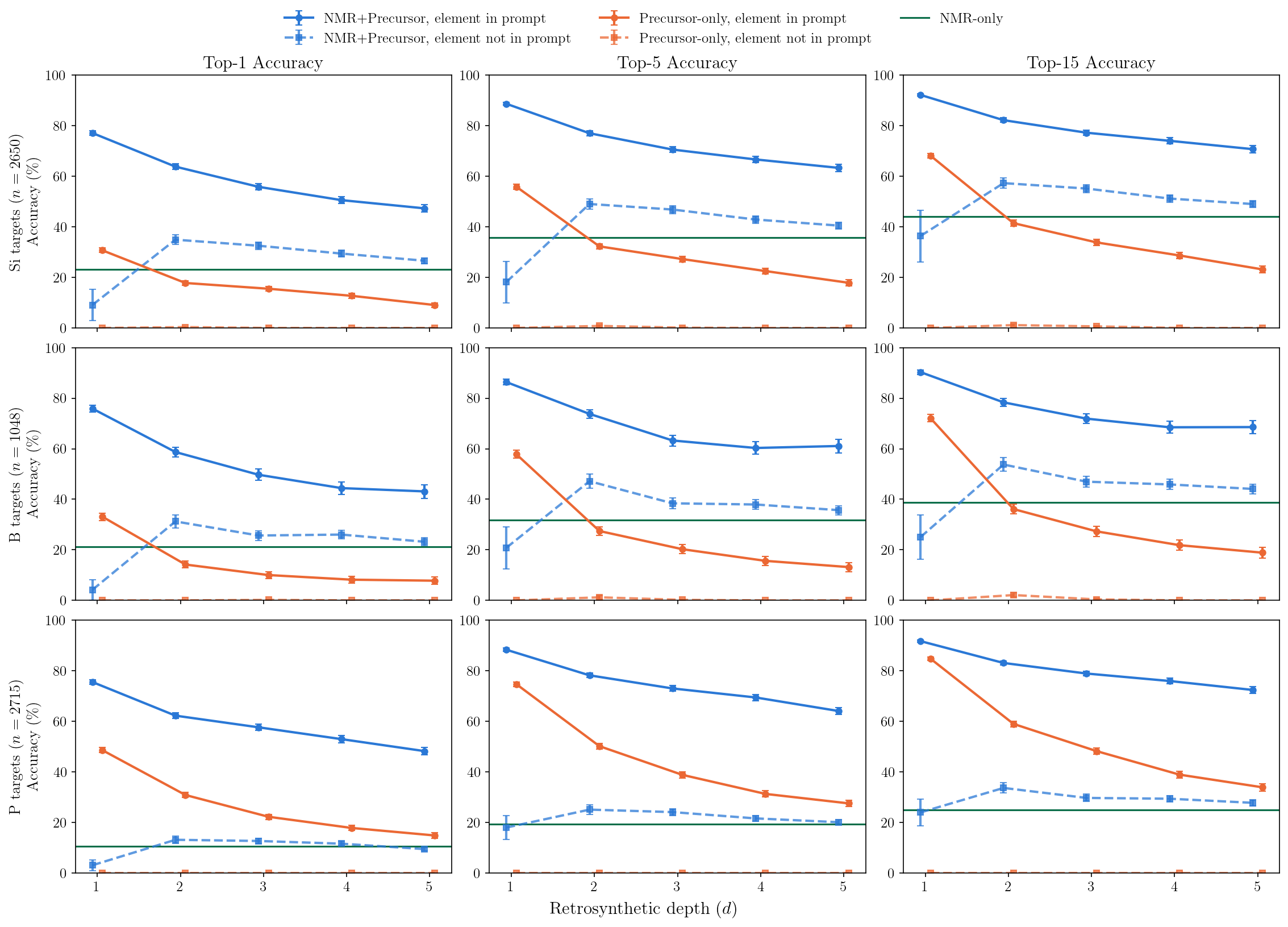}
    \caption{Top-$N$ accuracy ($N=1,5,15$, left to right) as a function of retrosynthetic depth for target molecules containing Si (top row), B (middle row), and P (bottom row), for the NMR+Precursor (blue), Precursor-only (orange), and NMR-only (green) models. Results are separated by whether at least one precursor in the prompt contains that element (solid lines, circles) or not (dashed lines, squares). Error bars show the standard error of each accuracy, which depends on the number of prompts at that depth and prompt type. The green horizontal lines show the accuracy of the NMR-only model on the same molecules (Top-1, Top-5, and Top-15 accuracies of 23.2\%, 35.8\%, and 44.1\% for Si; 21.2\%, 31.8\%, and 38.7\% for B; and 10.5\%, 19.2\%, and 25.0\% for P). All results are for the 2,650 Si-, 1,048 B-, and 2,715 P-containing molecules of the full test set, with one randomly selected prompt per molecule at each retrosynthetic depth when possible, excluding prompts in which the target itself appears as a precursor. The Top-15 panels are shown in the main text (Figure~4).}
    \label{SI_fig:rare_elements_full}
\end{figure}

Figure~\ref{SI_fig:rare_elements_full} shows the Top-1, Top-5, and Top-15 accuracies for target molecules containing Si, B, and P, separated by whether the prompt contains the element.

\FloatBarrier  

\section*{References}
\bibliography{references}